\documentclass{aastex}
\usepackage{url}

\RequirePackage{color}

\begin{document}

\title{The visual binary AG Tri in $\beta$ Pictoris Association: can a debris disc  cause very different rotation periods of its components?}
\slugcomment{Not to appear in Nonlearned J., 45.}
\shorttitle{The visual binary AG Tri}
\shortauthors{Messina et al.}

\author{Sergio Messina\altaffilmark{1}} 
\affil{INAF- Catania Astrophysical Observatory, via S.Sofia, 78 I-95123 Catania, Italy}
\and
\author{Miguel Muro Serrano\altaffilmark{2}}
\affil{Zeta UMa Observatory, Madrid, Spain}
\and
\author{Svetlana Artemenko\altaffilmark{3}}
\affil{Research Institute Crimean Astrophysical Observatory, 298409, Nauchny, Crimea}
\and
\author{John I. Bailey, III \altaffilmark{4}}
\affil{Astronomy Department, University of Michigan, USA}
\and
\author{Alexander Savushkin\altaffilmark{3}}
\affil{Research Institute Crimean Astrophysical Observatory, 298409, Nauchny, Crimea}
\and
\author{Robert H. Nelson\altaffilmark{5}}
\affil{Sylvester Robotic Observatory, 1393 Garvin Street, Prince George, BC, Canada}



\begin{abstract}
We measure the photometric rotation periods of the components of multiple systems in young stellar associations
to investigate the causes of the observed rotation period dispersion. We present the case of the wide
binary \object{AG Tri} in the 23-Myr young \object{$\beta$ Pictoris Association} consisting of K4 + M1 dwarfs. 
Our multi-band, multi-season photometric monitoring allowed us to measure the rotation periods of both components P$_{\rm A}$ = 12.4\,d and P$_{\rm B}$ = 4.66\,d, to detect a prominent magnetic activity
in the photosphere, likely responsible for the measured radial velocity variations, and for the first time, a flare event on the M1 component AG Tri B. We investigate either the possibility
that the faster rotating component may have suffered an enhanced primordial disc dispersal, starting its PMS spin-up
earlier than the slower rotating component, or the possibility that the formation of a debris disc may have prevented AG Tri A from 
gaining part of the angular momentum from the accreting disc. 
\end{abstract}


\section{Introduction}

Low-mass stars   (M $\la$ 1.2\,M$_\odot$ and spectral type later than about F5) \rm in young Open Clusters   ($\la$ 500 Myr) \rm  and Associations   ($\la$ 100 Myr) \rm exhibit a distribution of the rotation periods. Generally,
mass-period diagrams display an upper bound in the distribution whose bluer-part (consisting of stars,   starting from the F spectral type, \rm  that first settle on the ZAMS)
moves   progressively \rm toward longer rotation periods and with a decreasing 
dispersion as far as the stellar age increases (see, e.g., \citealt{Mamajek08}). 
By an age of about 0.5\,Gyr, F-G-K stars exhibit an almost one-to-one correspondence 
between rotation period and mass, as in the case of \object{Hyades} (\citealt{Delorme11}), \object{Praesepe} (\citealt{Douglas14}) and \object{Coma Berenicis} (\citealt{CollierCameron09})
open clusters. Such an univocal dependence is currently exploited for gyro-chronological estimate of stellar age (see, e.g., 
\citealt{Mamajek08}).\\
Within the same stellar cluster, the distribution of rotation periods depends on mass and, for any mass bin, it depends on the 
initial rotation period and on the angular momentum evolution that can vary from star to star. A key role in the early pre-main-sequence rotational evolution 
is played by the timescale of the star-disc locking, that is the magnetic coupling 
between the accreting disc and the star's external envelope   (see, e.g., \citealt{Camenzind90}; \citealt{Koenigl91}; \citealt{Shu94}). 
The shorter the disc lifetime, and its locking to the star,   the \rm earlier the star starts spinning up to become a fast rotator, owing to stellar radius contraction.\\
To investigate the effect played by the initial rotation period and by the disc-locking timescale on the observed rotation period distribution,
components of multiple stellar systems are particularly suited. In such systems, initial chemical composition, age, and in a few cases also the masses,
are equal, and any difference in the components' rotation periods must be attributed to only differences in the initial values of rotation periods 
and timescale of disc-locking.
Three such systems,  \object{BD -21 1074} (\citealt{Messina14}), \object{HIP 10680}/\object{HIP 10679} (\citealt{Messina15}) in $\beta$ Pictoris,  and \object{TYC 9300 891 1}AB/\object{TYC 9300 529 1}
(\citealt{Messina16}) in the Octans Association, were already investigated. In the first two cases, the components exhibit significant rotation period differences that
can be attributed to different disc lifetimes. In the latter case, the rotation period difference is relatively small   ($\sim$ 16\%) indicating that all components had similar or slightly different, at most, initial rotation periods and disc lifetimes. \rm \\
We try to investigate the hypothesis according to which when a star   with a stellar companion on a wide orbit has also a sufficiently close-in \rm companion of equal or lower mass, 
the gravitational effects of it on the primordial disc enhance its dispersal and shorten its life time and, consequently, the duration 
of the disc-locking phase. This circumstance is expected to determine a significant difference between the rotation periods of 
the two main components of the stellar system, where the more distant component will be found to rotate slower than the other component with 
the closer companion. This scenario was proposed for \object{BD -21 1074}. In the case of     \object{TYC 9300 891}\,1AB/ \object{TYC 9300 529 1}, \rm the companion at about 160\,AU seems to be too distant to significantly affect the disc lifetime and the rotation period difference is found to be relatively small.\\
On a different approach, if we observe a close visual binary and find that one component has a rotation period
significantly shorter than the other, we rise the suspect that the faster rotating component must have an
undiscovered yet companion that has enhanced its primordial disc dispersal making   the star rotating \rm  faster than the more
distant component.\\
On the other hand, the formation of a massive debris disc can lock part of the angular momentum away from the star, whereas
in the case of no formation the primordial disc accretes onto the star that receives more angular momentum with respect to the hosting planet/disc star.
This scenario was proposed for  \object{HIP 10680}/ \object{HIP 10679} and is now proposed for the currently analyzed target  \object{AG Tri}.
\\
Considering this controversial behavior, a numerous sample of such stellar systems is demanded to address on a statistical basis
the possible causes of rotation period differences.

Indeed, one of such systems is AG Tri that consists of two components of about 0.85 and 0.45\,M$_{\odot}$ and very different rotation
periods. We suspect that the disc of the A component, which has a rotation period twice longer than the B component, has   held \rm part of the system angular momentum preventing the central star to spin up as did, on the contrary, the disc-less companion.

\section{Literature information}
		 \object{AG Tri} is  a wide visual binary consisting of two physically bound Pre-Main-Sequence stars at a distance d = 42.3\,pc to the Sun (\citealt{Rodriguez12}) and separated by $\rho$ = 22$^{\prime\prime}$ (\citealt{Mason01}),   which corresponds to about 930\,AU. \rm The components are reported in the literatures as K7/8 and M0 stars (\citealt{Rodriguez12}). However, our analysis shows
the correct spectral types are K4 and M1, respectively   (see Sect.\,5). \rm 
This system was first proposed as member of the $\beta$ Pictoris Associations by \citet{Song03}. This membership was subsequently 
confirmed by \citet{Lepine09}, and considered as bona fide member of this Association in the subsequent studies (see, e.g.,  \citealt{Malo13}, \citealt{Malo14}). 
  Given \rm their young age of 23$\pm$3 Myr (\citealt{Mamajek14}), both components have well detectable lithium lines. The measured Li equivalent widths are in the range 215--248\,m\AA\,\,for the A component and 110--130\,m\AA\,\,for the B component (\citealt{Mentuch08}; \citealt{daSilva09}; \citealt{Xing12}; \citealt{Malo14}).
The literature projected rotational velocities are $v \sin{i}_A$ =  5\,km\,s$^{-1}$ (\citealt{Cutispoto00}); $v \sin{i}_A$ =  4.7\,km\,s$^{-1}$ 
and $v \sin{i}_B$ =  5\,km\,s$^{-1}$ (\citealt{Bailey12}).\\
The presence of a debris disc around the A component was first detected by \citet{Rebull08} using the MIPS (Multiband Imaging Photometer for Spitzer) instrument
onboard the Spitzer Space Telescope
and, recently, by \citet{Riviere-Marichalar14} from Herschel Space Observatory observations.
The rotation period P = 13.6828\,d of the unresolved system was measured by \citet{Norton07}  from SuperWASP data collected during the 2004 run. Subsequently,  \citet{Messina11}  reported P = 12.5\,d, attributed to the brighter component A,  analyzing the complete 2004--2008 timeseries. However, these observations could not spatially resolve the two 
components and the rotation period of the B component remained unknown.

\begin{figure*}
\begin{centering}
\includegraphics[width=70mm,height=80mm,angle=0,trim= 0 0 0 0]{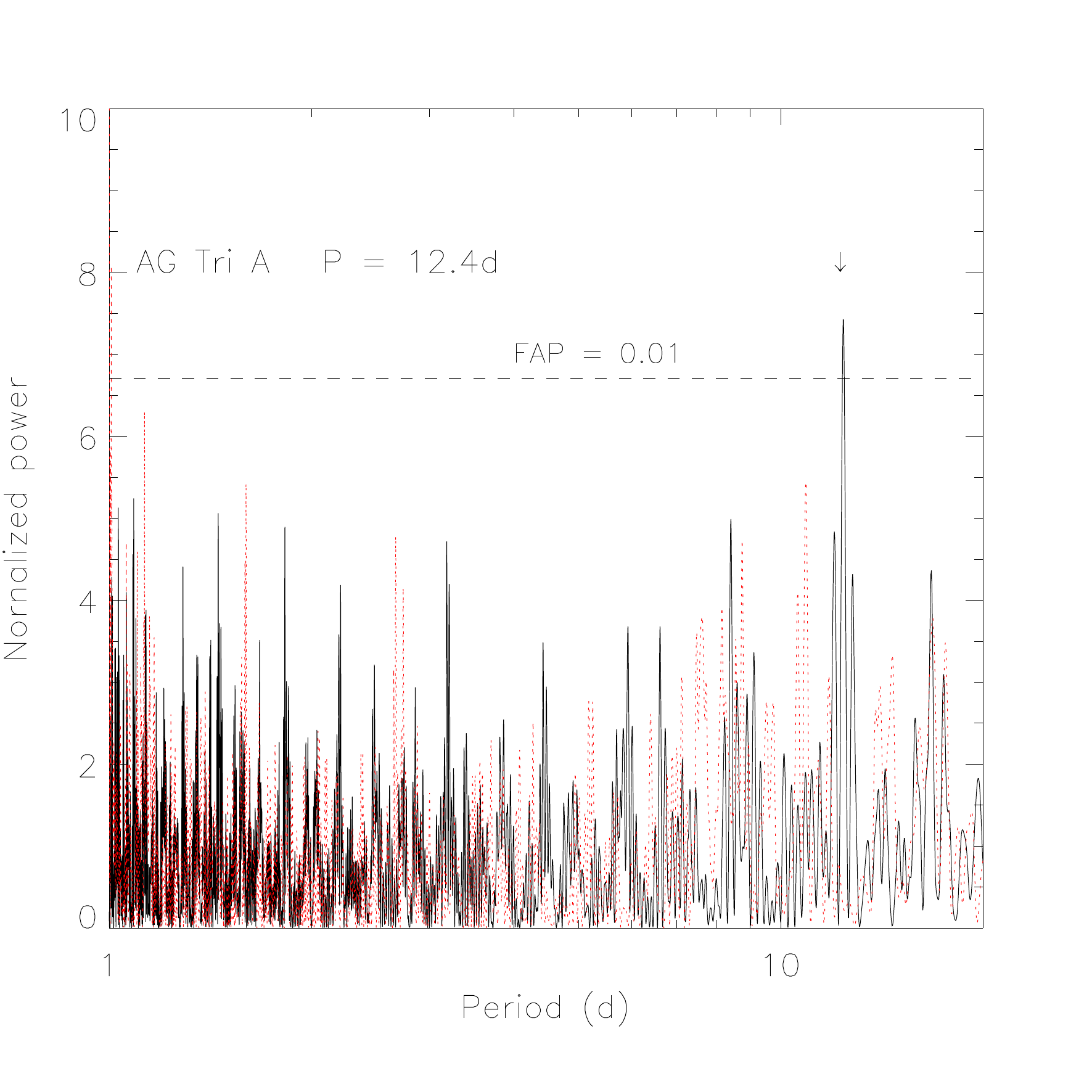}
\includegraphics[width=70mm,height=80mm,angle=0,trim= 0 0 0 0]{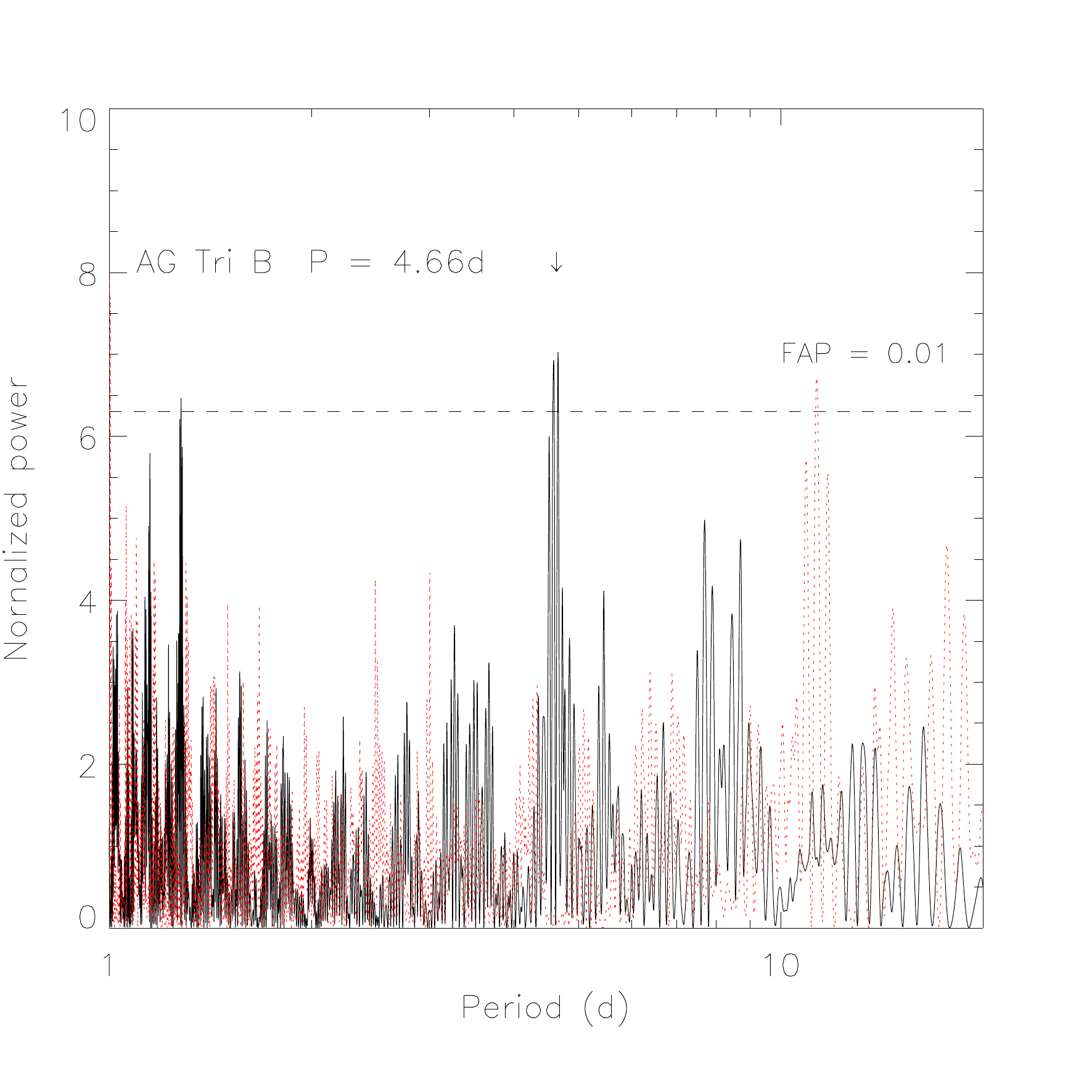}
\caption{Lomb-Scargle periodograms of the complete V-magnitude time series of AG Tri A (left panel) and of AG Tri B (right panel)
collected at the  Zeta UMa and CrAO Observatories. The dotted line represents the spectral window function, whereas the horizontal dashed line indicates 
the power level corresponding to a FAP = 1\%. The power peak corresponding to the rotation periods P = 12.4\,d of AG Tri A and  P = 4.66\,d  of AG Tri B are  indicated by  downward arrows.}
\label{periodogram_mizar}
\end{centering}
\end{figure*}

\begin{figure*}
\begin{centering}
\includegraphics[width=80mm,height=120mm,angle=0,trim= 0 0 0 0]{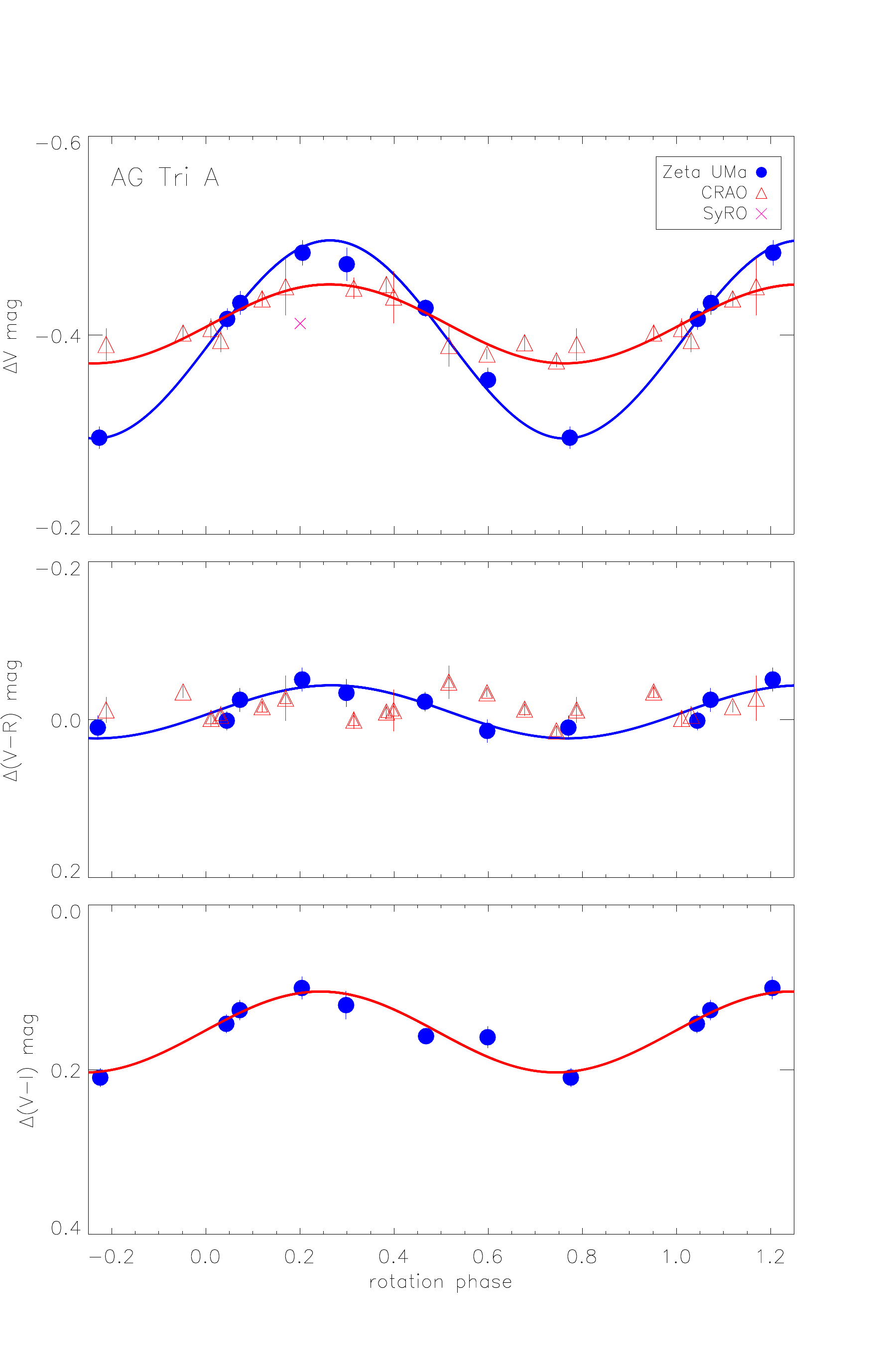}
\includegraphics[width=80mm,height=120mm,angle=0,trim= 0 0 0 0]{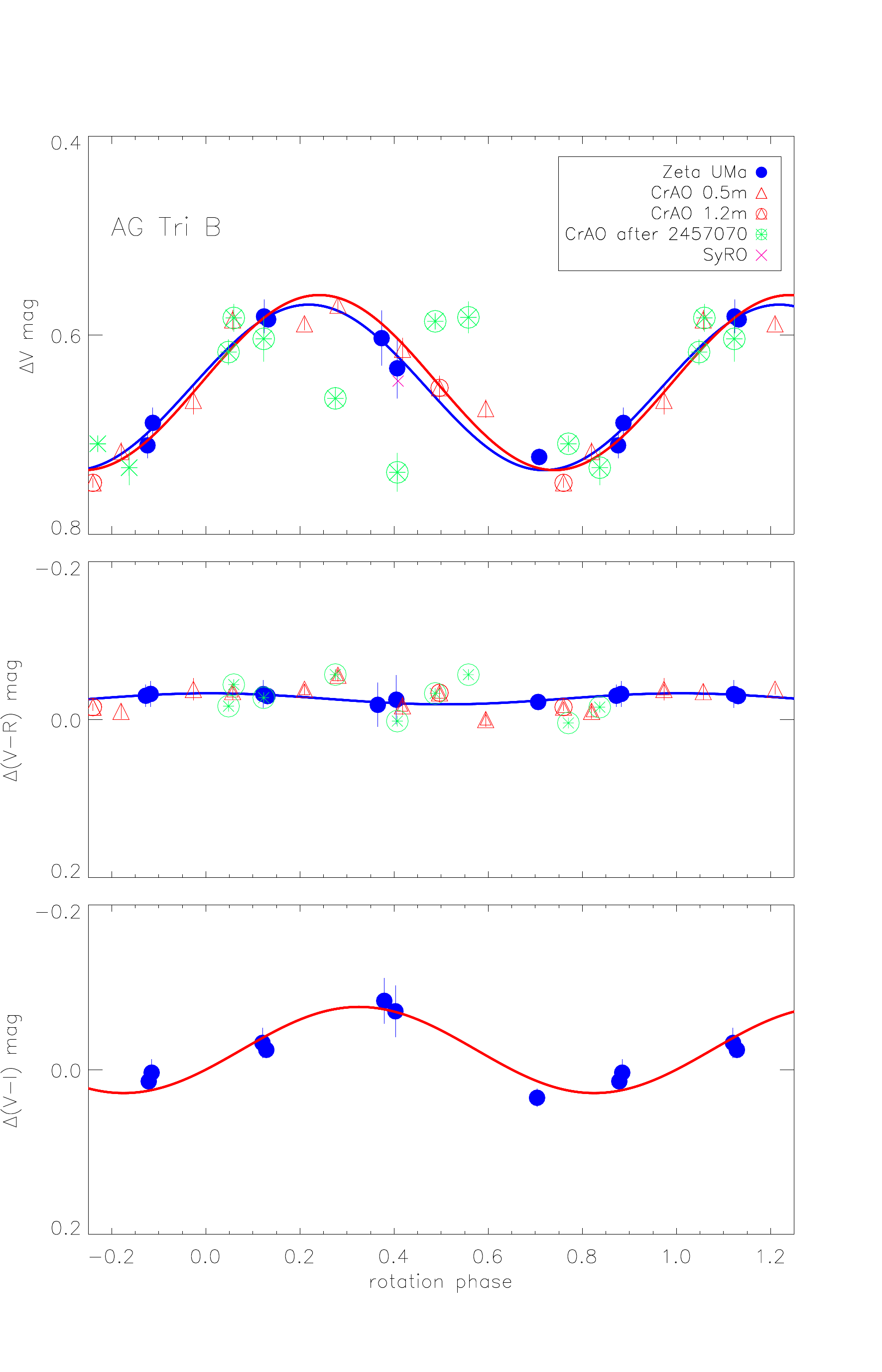}
\caption{Differential V-band light curves and V$-$R and V$-$I color curves of AG Tri A (left panel) and of AG Tri B (right panel) obtained with the data
collected at the  Zeta UMa (blue bulltes) and CrAO (red triangles and circled green asterisks) observatories, after adding a magnitude offset.
Rotation phases of AG Tri A are computed using the rotation period P = 12.4\,d, whereas for AG Tri B  using the rotation period P = 4.66\,d.}
\label{mizar_lightcurve}
\end{centering}
\end{figure*}

\section{Photometric observations}
To measure the rotation periods of the resolved components,  we planned our own photometric monitoring which was carried out
in two different seasons at two different Observatories.

\subsection{Zeta UMa Observatory}
We carried out multiband photometric observations of AG Tri at the Zeta UMa Observatory (709\,m a.s.l, Madrid, Spain). 
The observations were collected by a 130mm f/1.7 Takahashi refractor equipped with a cooled QHY9 camera and with a 
set of  Johnson-Cousins V, R, and I  filters. We observed a field of about 80$^\prime$ $\times$ 60$^\prime$ 
centered on AG Tri. The angular resolution at the focal plane was 1.50$^{\prime\prime}$/pixel.
The observations were carried out for a total of 7 nights from December 7, 2013 to March 09, 2014. Bad weather conditions prevented us from
collecting more data. We could collect a total of 317 frames in V,  168 in R, and 122 in I filter. Two different integration times were used for each filter. 
Longer exposures  (80s, 50s, 60s in V, R, and I filters) were used to achieve   a  S/N ratio better than 100 \rm  for the fainter component AG Tri B, whereas
shorter exposures (15s, 10s, 20s in V, R, and I filters) were used to avoid saturation of the brighter component AG Tri A. 
In all frames the two components of AG Tri were spatially well resolved.\\
The data reduction was carried out using the package DAOPHOT  within IRAF\footnote{IRAF is distributed by the National Optical Astronomy Observatory, which 
is operated by the Association of the Universities for Research in Astronomy, inc. (AURA) under 
cooperative agreement with the National Science Foundation.}. After bias subtraction and flat fielding, we extracted the magnitudes of all stars detected in each frame using a set of different apertures. Then, we selected the aperture giving the best photometric precision of our targets and comparison stars. 
We identified six stars close to AG Tri that were found to be non-variable during the whole period of our observations and suited to build an ensemble comparison star 
to get differential magnitudes of our targets. Magnitudes collected on the same night (generally consisting of a sequence of 12 exposures lasting no longer than half an hour\footnote{Only during the first observation night we collected 70 consecutive frames in the V filter for a total of about 2.5\,hr.}) 
were averaged to obtain one data point and its standard deviation that we consider 
as the photometric precision we could achieve. In the case of AG Tri A we obtained the following photometric precisions: $\sigma_{\rm V}$ = 0.012\,mag, 
$\sigma_{\rm R}$ = 0.008\,mag, and $\sigma_{\rm I}$ = 0.004\,mag. 
In the case of AG Tri B we obtained the following photometric precisions: $\sigma_{\rm V}$ = 0.020\,mag, $\sigma_{\rm R}$ = 0.010\,mag, and $\sigma_{\rm I}$ = 0.012\,mag.
 We found that the average standard deviation for the comparison stars (C$_i$) with respect to the ensemble star  was $\sigma_{Ci-Ens}$ = 0.006\,mag   in the V filter, slightly better in the R and I filters. \rm

\begin{figure}
\includegraphics[width=50mm,height=70mm,angle=90,trim= 0 0 0 0]{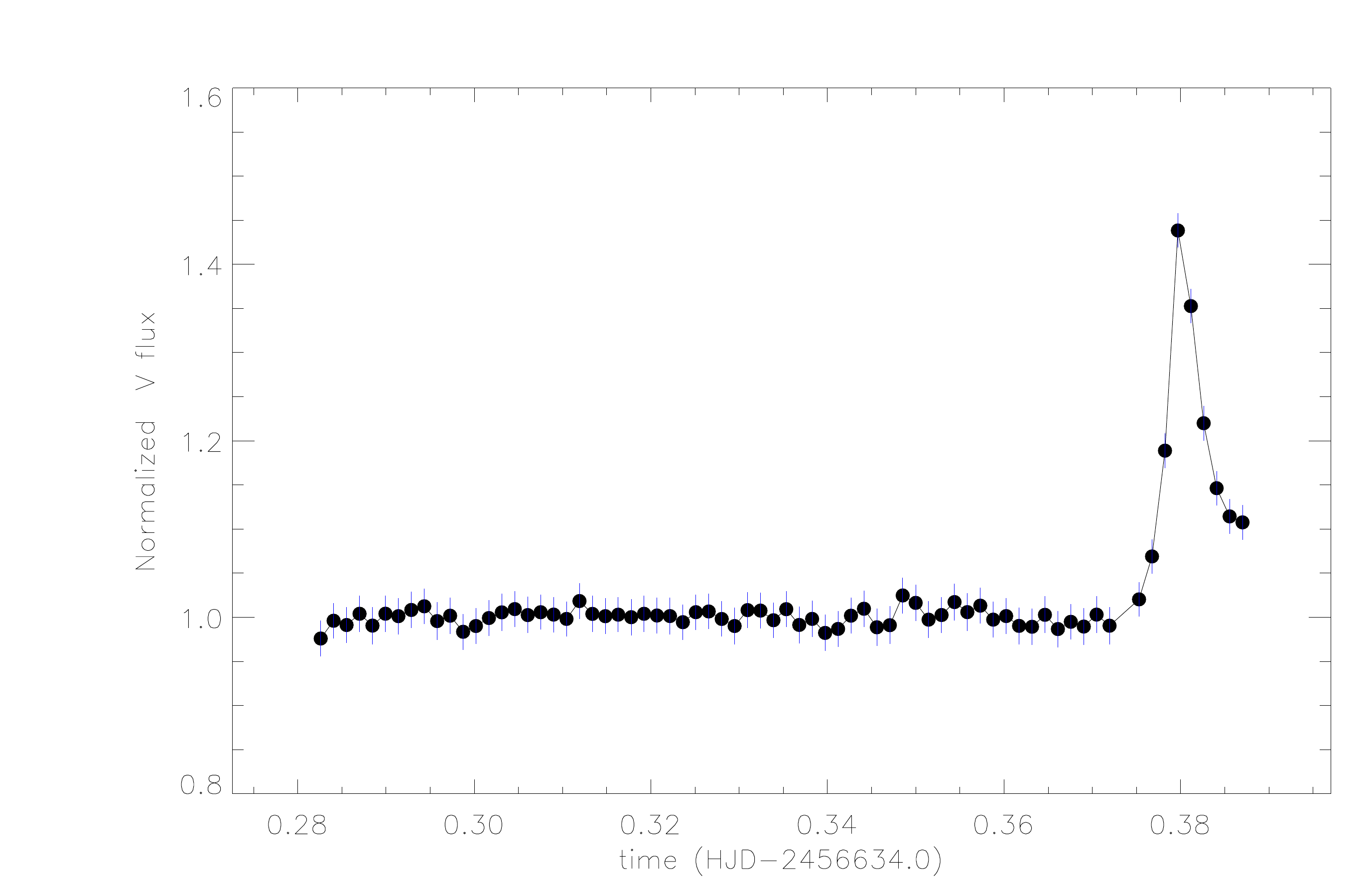}
\caption{Flare event detected in the V band during the first observation night at the Zeta UMa Observatory.}
\label{flare_V}
\end{figure}

\subsection{CrAO observations}

The system AG Tri was observed from October 29, 2014 till March 9, 2015 for a total of 19 nights at the Crimean Astrophysical Observatory (CrAO, 600\,m a.s.l., $+$44$^{\circ}\,43^{\prime}\,37^{\prime\prime}$ and 34$^{\circ}\,01^{\prime}\,02^{\prime\prime}$E, Nauchny, Crimea). The observations were collected   by two different telescopes. In 17 nights the targets were observed  by   the  0.5m \rm Maksutov telescope equipped with an Apogee Alta U6 CCD camera (1024$\times$1024 pixels, with a field of view 12.2$^{\prime}$ $\times$12.2$^{\prime}$, and angular resolution 0.71$^{\prime\prime}$/pixel), with V and R Johnson filters. \\
In only two nights (Nov 11 and Dec 01) the targets were observed by the 1.25m Ritchey-Chr\'etien telescope equipped with FLI ProLine PL230 CCD camera (2048$\times$2048 pixels, with field of view 10.9$^{\prime}$$\times$10.9$^{\prime}$, and angular resolution  0.32$^{\prime\prime}$/pixel) with the same filters. \\
We could collect 103 frames in the V filter  and 102 in the R filter using 30\,s and 15\,s exposures, respectively, using the 0.5m Maksutov telescope. We could collect 10 frames in the V filter  and 10 in the R filter, using the 1.25m Ritchey-Chr\'etien telescope, and adopting the same exposure times. \rm Bias subtraction, flat field correction, and magnitude extraction
were done using the IRAF routines, as described in the previous Section\,3.1. 
Each telescope pointing consisted of five consecutive exposures that were subsequently averaged to obtain one average magnitude
for each telescope pointing. The average standard deviation associated to the averaged magnitude is $\sigma_{\rm V}$ = 0.011 mag   ($\sigma_{\rm V}$ = 0.009 mag for the 1.2m telescope) \rm and 
$\sigma_{\rm R}$ = 0.010 mag   ($\sigma_{\rm R}$ = 0.006 mag for the 1.2m telescope), \rm which we consider the average photometric precision we could achieve.
Owing to the smaller FoV of the two telescopes with respect to the Zeta UMa telescope, we identified three stars suitable to serve as comparison stars, and that were used to build a new ensemble comparison star.\\
To combine data collected from these two different telescopes into a single time series, we phased the V magnitudes and V$-$R colors of the more numerous data set (i.e., the data collected with the 0.5m telescope) using the known rotation period of AG Tri A  taken from the literature (\citealt{Messina11}). \rm Then, we added a magnitude offset to the data collected with the 1.2m telescope in such a way to minimize their phase dispersion with respect to the other data set. The same magnitude offsets were also applied to AG Tri B data collected with the 1.2m telescope. \rm 

\subsection{Sylvester Robotic Observatory}
A first attempt by us to observe AG Tri was made on December 12, 2012 at the Sylvester Robotic Observatory (SyRO; Prince George, BC, Canada).
 We used a 33cm f/4.5 Newtonian telescope mounted on a Paramount ME,
and equipped with a SBIG ST-10XME CCD camera (2184$\times$1472 pixels)  with a field of view 34.4$^{\prime}$ $\times$23.2$^{\prime}$ 
and angular resolution 0.95$^{\prime\prime}$/pixel. We could collect a series of 22 frames in the V band using 120s exposure time.
Bias subtraction, flat field correction, and magnitude extraction
were done using the IRAF routines, as described in the previous Section 3.1,   using the same set of comparison stars used for
the Zeta UMa data analysis. However, it was not used in the subsequent analysis for the rotation period search. \rm

 \section{Periodogram analysis}
 \subsection{AG Tri A}
We used the Lomb-Scargle (LS; \citealt{Scargle82}) and CLEAN (\citealt{Roberts87})  periodogram analyses to measure the rotation period of  AG Tri A. 
In the left panel of Fig.\,\ref{periodogram_mizar}, we show the LS periodogram obtained with the complete V-band data collected at the Zeta UMa and CrAO observatories.
Since we adopted for the CrAO data a different ensemble comparison, a magnitude offset was added to the CrAO magnitude time series to be comparable (same average magnitude) with the Zeta UMa magnitude time series.   This operation removes any possible  intrinsic variation (owing to the magnetic activity evolution) of the average magnitude between the two datasets collected at about 1 yr of distance, as well as the \rm  effects arising from the use of two different instruments. \rm 
 The highest power peak is found at P = 12.4$\pm$0.5\,d and with False Alarm Probability FAP$<$ 1\%,
 that is the probability that a power peak of that height simply arises from Gaussian noise in the data. 
 The FAP was estimated using a Monte-Carlo method, i.e., by generating 1000 artificial light curves obtained from the real one, keeping the date but scrambling the magnitude values (see e.g. \citealt{Herbst02}).
The uncertainty of the rotation period is computed following the prescription of \citet{Lamm04}. Similar results were obtained with the CLEAN algorithm. This 
rotation period determination is in agreement with the earlier determination by \cite{Messina11} on the unresolved system. Indeed, now we are sure that the earlier determination refereed to the brighter A component.

In the left panels of Fig.\,\ref{mizar_lightcurve} we plot the differential V-band and the V$-$R and V$-$I color curves phased with the rotation period
P = 12.4\,d for AG Tri A using blue bullets for data collected at Zeta UMa observatory in the 2013/2014 season and red triangles for data collected at CrAO observatory 
in the 2014/2015 season.  AG Tri A exhibits  a significant variation of the V-band light curve amplitude: from $\Delta$V = 0.20 mag in the first season to $\Delta$V = 0.08 mag in the second season.
The V$-$R color variation was $\Delta$(V$-$R) = 0.07\,mag, whereas in the second season it appeared quite scattered. Finally, in the first season we measured a color variation $\Delta$(V$-$I) = 0.09 mag.\\

 \subsection{AG Tri B}
 We used the same Lomb-Scargle and CLEAN  periodogram analyses to measure the rotation period of  AG Tri B.
 In the right panel of Fig.\,\ref{periodogram_mizar}, we show the LS periodogram obtained with the complete V-band data collected at the Zeta UMa and CrAO observatories.
We added a magnitude offset to the CrAO magnitude time series to be comparable (same average magnitude) with the Zeta UMa magnitude time series.
 The highest power peak is found at P = 4.66$\pm$0.05\,d and with False Alarm Probability FAP$<$ 1\%. The secondary peak at P = 1.27\,d is a beat of the rotation period.
 
As in the previous case, in the right panels of Fig.\,\ref{mizar_lightcurve}, we plot the differential V-band and the V$-$R and V$-$I color curves phased with the rotation period
P = 4.66\,d for AG Tri  B using blue bullets for data collected at Zeta UMa observatory in the 2013/2014 season and red triangles for data collected at CrAO observatory 
in the first part of the 2014/2015 season. We used green circled asterisks to indicate the CrAO data collected in the second part of the 2014/2015 season.
Whereas the Zeta UMa data and the first part of the CrAO data time series exhibit same amplitude, phase of minimum , and shape with the following light curve amplitudes: 
$\Delta$V = 0.16 mag, $\Delta$(V$-$R) = 0.02\,mag, and  $\Delta$(V$-$I) = 0.08 mag, in the second part of the CrAO data time series (from HJD 2457072 till 2457091), 
the light curve exhibits a double minimum, a smaller amplitude $\Delta$V = 0.15 mag, and a quite scattered $\Delta($V$-$R) color curve with no evident rotational modulation and magnitude dispersion of 0.014 mag.  \rm The double minimum likely 
arises from the presence of two active regions on opposite stellar hemispheres.

We note that for both components AG Tri A and B, the V-band light curve and the color curves are all correlated in phase, when the star
gets fainter it also gets redder. This behavior is consistent with the presence of active regions dominated by only cool spots or only hot spots.\\
Interestingly, during the first observation night, when data were collected for 2.5 consecutive hours in the V filter, we detected on AG Tri B a flare
event, measuring an increase of flux up to 40\% of the quiescent flux   (see Fig.\,\ref{flare_V}). \rm Unfortunately, observations were stopped before the end of this event.
This is the first flare detection on this component ever reported in the literature. This event occurred at rotation phase $\phi$ = 0.43, which corresponds to
middle way between the light curve maximum and minimum.
 

\section{Spectral parameters}

For the spectral parameters determination of both components, we made use of the spectra time series presented by Bailey et al. (2012). These spectra
 were obtained using the cross-dispersed infrared echelle spectrograph NIRSPEC (\citealt{McLean98}) on the W. M. Keck II telescope. 
 The observations were obtained over 12 observing runs between November 2004 and May 2009. All observations were obtained with the 3 pixel 
 (0.432$^{\prime\prime}$) slit in combination with the N7 blocking filter, and approximate echelle angle 62.65$^{\circ}$ and grating angle 35.50$^{\circ}$. 
 Analysis used NIRSPEC order \#33 which spans $\sim$270\,\AA, from 2.288 $\mu$m to 3.315\,$\mu$m at a resolving power of approximately 30.000. 
 The spectra were obtained in pairs at two locations along the slit which provided a nearly simultaneous measurement of sky 
 emission and detector bias for both images. Spectra have S/N ratios of $\sim$200; see the Bailey et al. (2012) paper for details. \rm

We refit the spectra of AG Tri A and B first reported in \citet{Bailey12} using an updated version of the YSAS pipeline (that will be detailed in a forthcoming paper). Two key changes are that it now uses the full PHOENIX synthetic (\citealt{Husser13})  spectrum grid interpolated to 50\,K in T$_{eff}$ and 0.1\,dex in $\log(g)$, [Fe/H], and [$\alpha$/Fe], and that a bug in the NASA IDL library function \texttt{lsf\_rotate} for rotational broadening has been found and corrected\footnote{The kernel size was not computed appropriately for small $v\sin{i}$ and could be also asymmetric}. We hold log(g) fixed at 4.7 for both stars and [Fe/H] and [$\alpha$/Fe] at the solar values. T$_{eff}$ and $v\sin{i}$ are not appreciably affected by this choice, perhaps as the spectral order contains predominately CO lines.    The micro-turbulent velocity of the two stars was assumed to be in the range $\sim$0.3--0.6\,km\,s$^{-1}$ (\citealt{Husser13}). \rm

For the 28 spectra of AG Tri A and B we find a T$_{eff}$ = $4334 \pm 27$\,K and T$_{eff}$ =  $ 3538 \pm 42$\,K, respectively, where the errors given are standard deviations of the best-fit values for the spectra. We note that the T$_{eff}$  we measure for AG Tri A corresponds to a K4 spectral type (\citealt{Pecaut13}), which is earlier than
either the K7 or K8 spectral types reported in the literature. This new K4 spectral type  is well consistent with the observed optical and infrared colors 
V$-$I = 1.32\,mag, V$-$K = 3.02\,mag, J$-$H = 0.63\,mag, and H$-$K = 0.16\,mag.
On the contrary, we note that the T$_{eff}$  we measure for AG Tri B corresponds to a M1 spectral type (\citealt{Pecaut13}), which is slightly later than the M0 reported in the literature. This new estimate is well consistent with the observed optical and infrared colors 
V$-$K = 4.52\,mag, J$-$H = 0.68\,mag, and H$-$K = 0.22\,mag.   Given the relatively small distance to the system, we considered negligible the interstellar reddening. \rm  As we will show, these new effective temperature measurements allow both components to be best fitted
by the same isochrone, within the uncertainties, as expected for coeval components of a physical binary.
For the projected rotational velocities we find  $v\sin{i}$ = $5.25 \pm 0.75$\,km\,s$^{-1}$ and $v\sin{i}$ = $7.46 \pm 0.75 $\,km\,s$^{-1}$ for AG Tri A and B, respectively.

\begin{figure}
\begin{minipage}{9cm}
\includegraphics[width=50mm,height=80mm,angle=90,trim= 0 0 0 0]{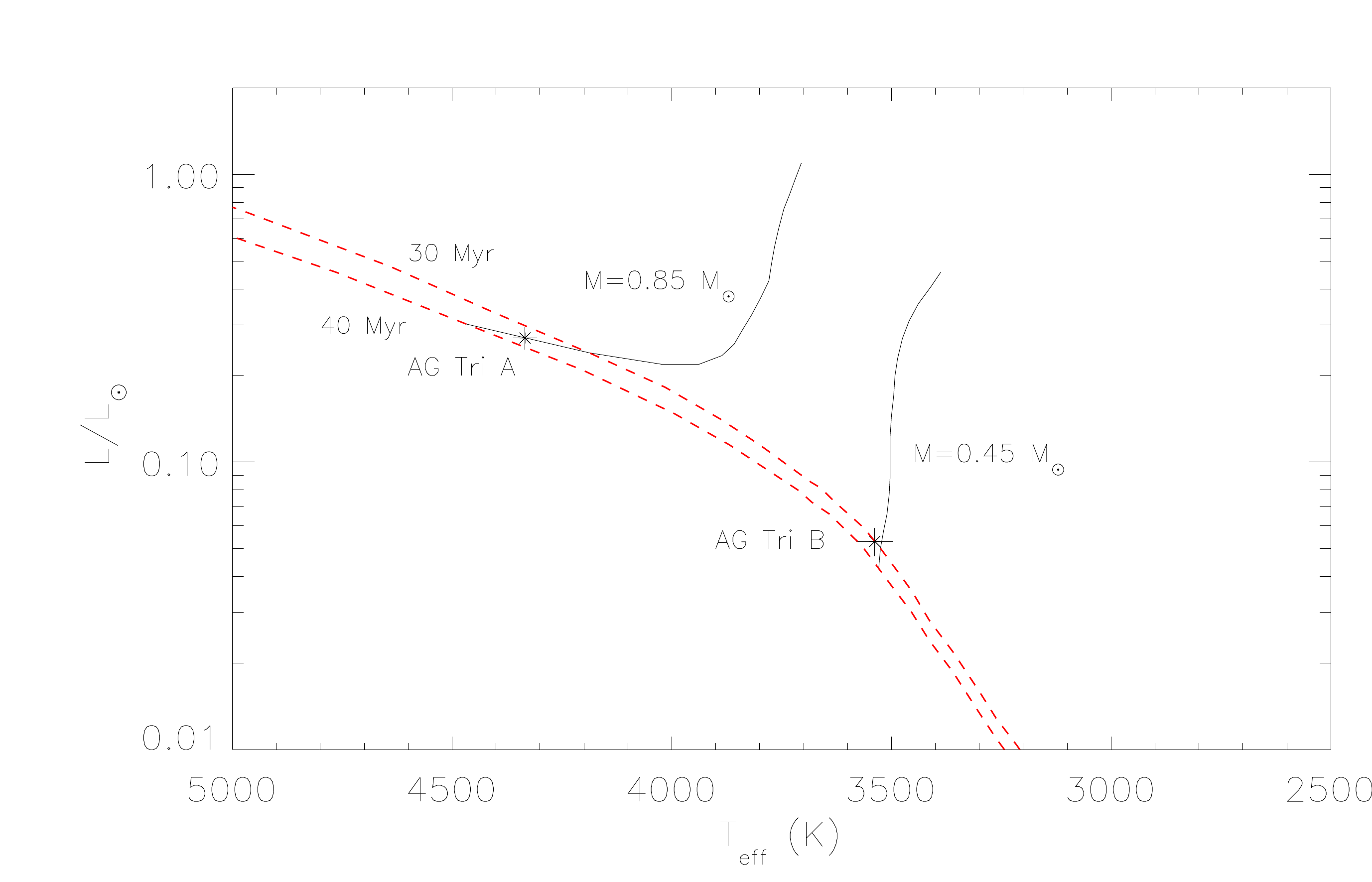} 
\includegraphics[width=50mm,height=80mm,angle=90,trim= 0 0 0 0]{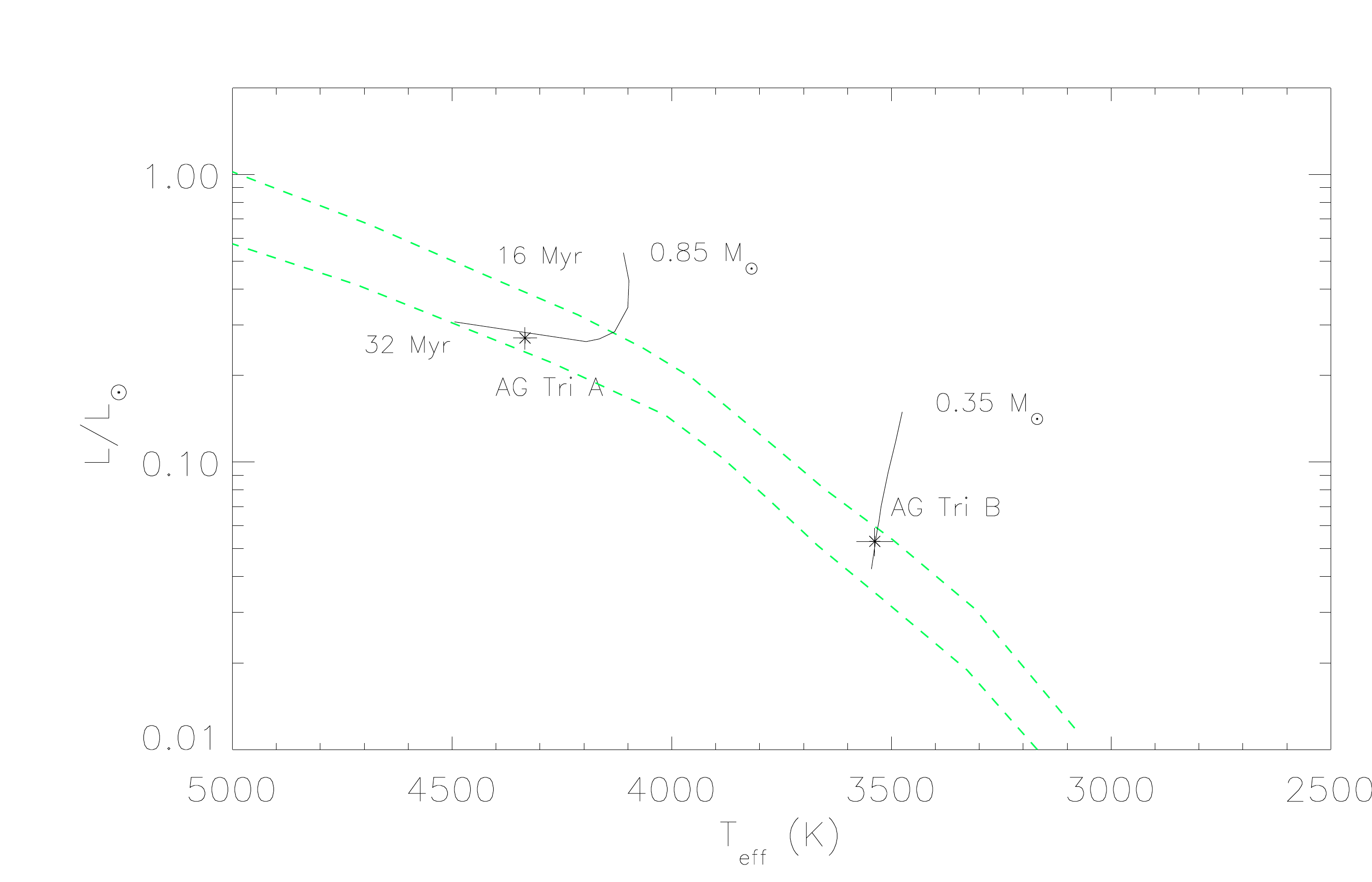} 
\includegraphics[width=50mm,height=80mm,angle=90,trim= 0 0 0 0]{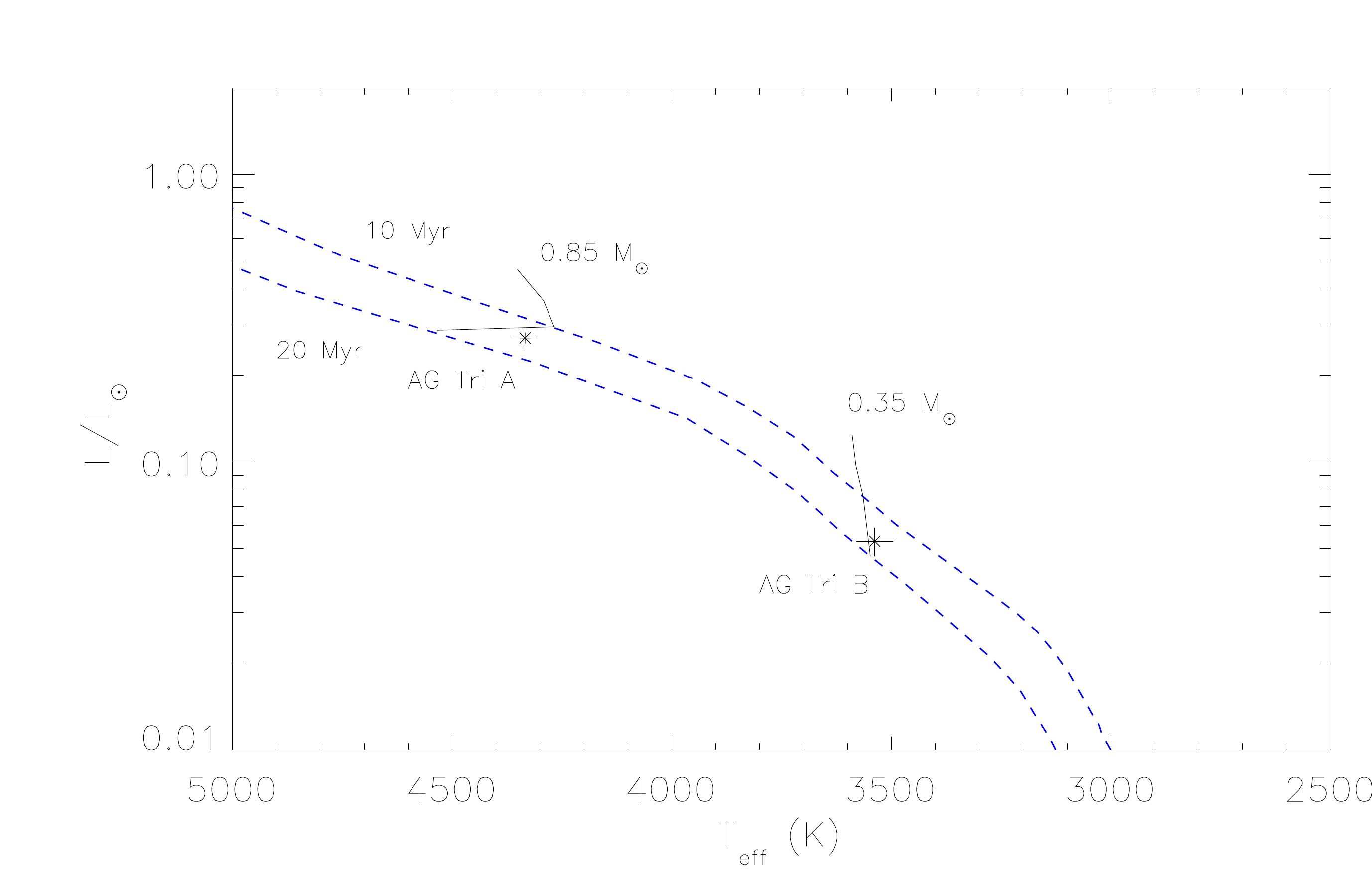} 
\end{minipage}
\caption{\label{hr}HR diagrams. Dashed lines are isochrones whereas the blue solid lines are the evolutionary mass tracks. Models are from \cite{Baraffe98} (top panel), \cite{Siess00} (middle panel), and \cite{D'Antona97} (bottom panel).}
\vspace{0cm}
\end{figure}

Using the visual magnitudes V$_{\rm A}$ = 10.10\,mag and V$_{\rm B}$ = 12.44\,mag,  distance d = 42.3\,pc,  bolometric correction BC$_{\rm VA}$ = $-$0.85\,mag  and BC$_{\rm VB}$ =  $-$1.58\,mag (\citealt{Pecaut13}), we derive the luminosities L$_{\rm A}$ = 0.27$\pm$0.03\,L$_\odot$ and L$_{\rm B}$ = 0.063$\pm$0.005\,L$_\odot$.
In Fig.\,\ref{hr}, we compare T$_{\rm eff}$ and luminosity  of both components with a set of isochrones taken   from \citet{Baraffe98}, \citet{Siess00}, and \citet{D'Antona97}, \rm for solar  metallicity to infer the age of each component and to see if the hypothesis of coevalness  is sustainable. 
 We see that the \citet{D'Antona97}  and \citet{Baraffe98} models produce the smallest age difference between the components with an estimated age of 35$\pm$5 Myr and 15$\pm$5\,Myr, respectively. The \citet{Siess00} model produces results with a slightly larger difference  with an estimated age of 24$\pm$8 Myr. According to these models the system AG Tri has an age of 24$\pm$10 Myr in agreement with the age of 23$\pm$3\,Myr estimated by \citet{Mamajek14}. All models provide the same mass M = 0.85$\pm$0.05\,M$_\odot$ for AG Tri A.   
For AG Tri B \rm we get a mass M = 0.35$\pm$0.05\,M$_\odot$ from Siess et al. and from D'Antona et al., whereas a sligtly larger value  M = 0.45$\pm$0.05\,M$_\odot$ from Baraffe et al. models.\\
 Using the measured effective temperatures T$_{\rm A}$ = 4334\,K and T$_{\rm B}$ = 3538\,K,
we derive the stellar radii R$_{\rm A}$ = 1.02$\pm$0.12\,R$_\odot$ and R$_{\rm B}$ = 0.68$\pm$0.09\,R$_\odot$. Combining rotation periods and projected rotational velocities, we can derive the inclinations of the rotation axes.   We found that \rm i$_B$ $\simeq$ 90$^{\circ}$. However, in the case of AG Tri A we derive $\sin{i} > 1$.   We must note that in our measurement of the projected rotational velocity we did not account for the 
macroturbulent-velocity-induced line broadening, which is of the order of $\sim$2.4\,km\,s$^{-1}$ for early K-type stars (\citealt{Gray84}). Therefore, corrected value should be  $v\sin{i}_{\rm A}$ = $4.66$\,km\,s$^{-1}$. However,  this correction is not enough. We must  assume that we are underestimating also the stellar radius. Actually, recent works by \citet{Jackson14} and by \citet{Somers15} show that the effect of starspots on PMS  stars is  to reduce the luminosity of the star while increasing its radius as $\sim$ 1/3 power of the unspotted photosphere (where the exponent is about 0.4 in our case). In our case,   assuming $i$ = 90$^{\circ}$, we find that the radius should be larger by $\sim$8\% to conceal rotation period and $v\sin{i}$. According to the mentioned work,
such an inflation would be possible if about 20\%
 of the photosphere of AG Tri A is covered by starspots. \rm

\section{Spot   modelling \rm}
\begin{figure*}
\begin{minipage}{10cm}
\includegraphics[width=70mm,height=100mm,angle=90,trim= 0 0 0 100]{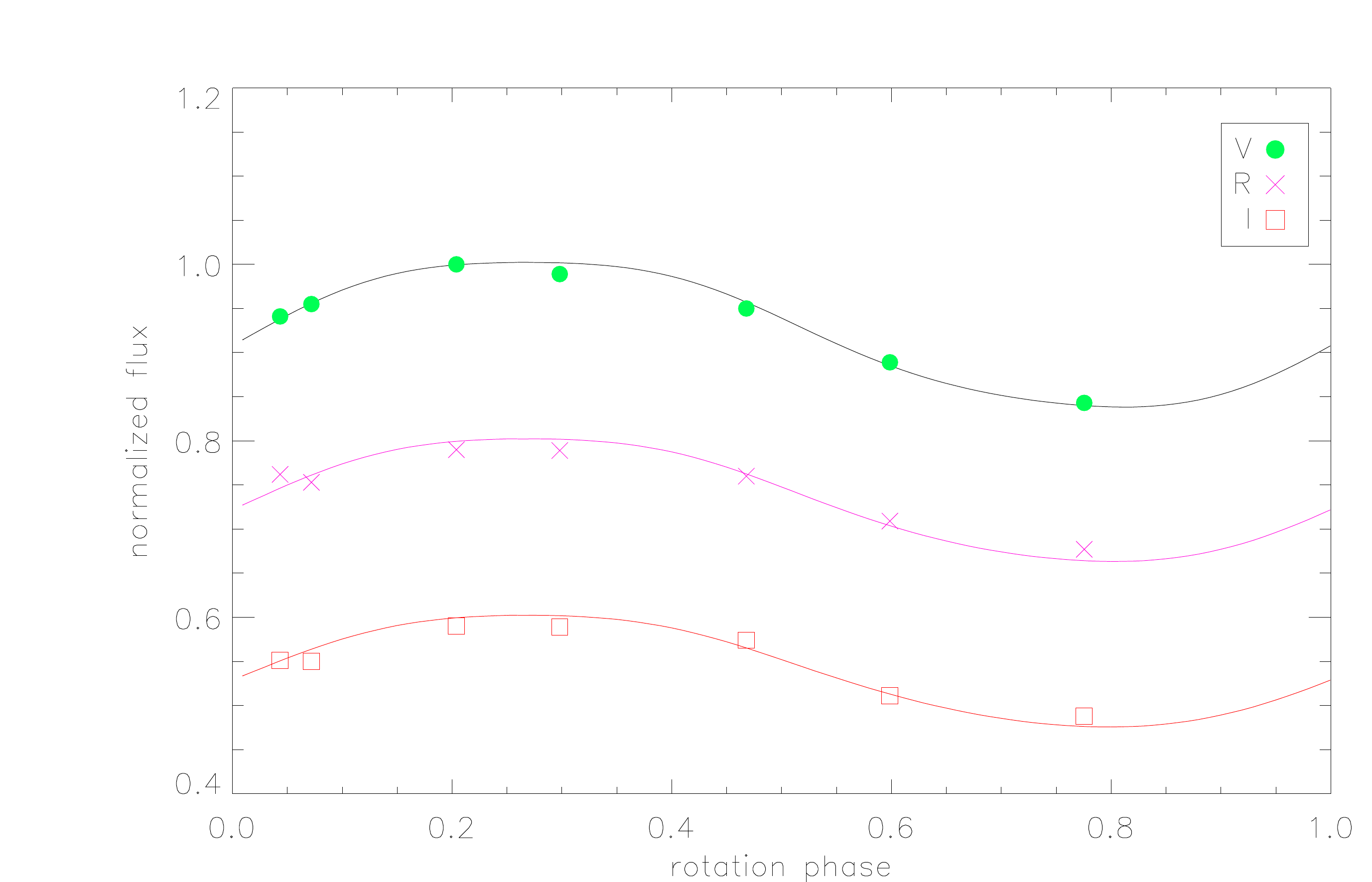}\\
\includegraphics[width=30mm,height=30mm,angle=0,trim= 0 0 0 0]{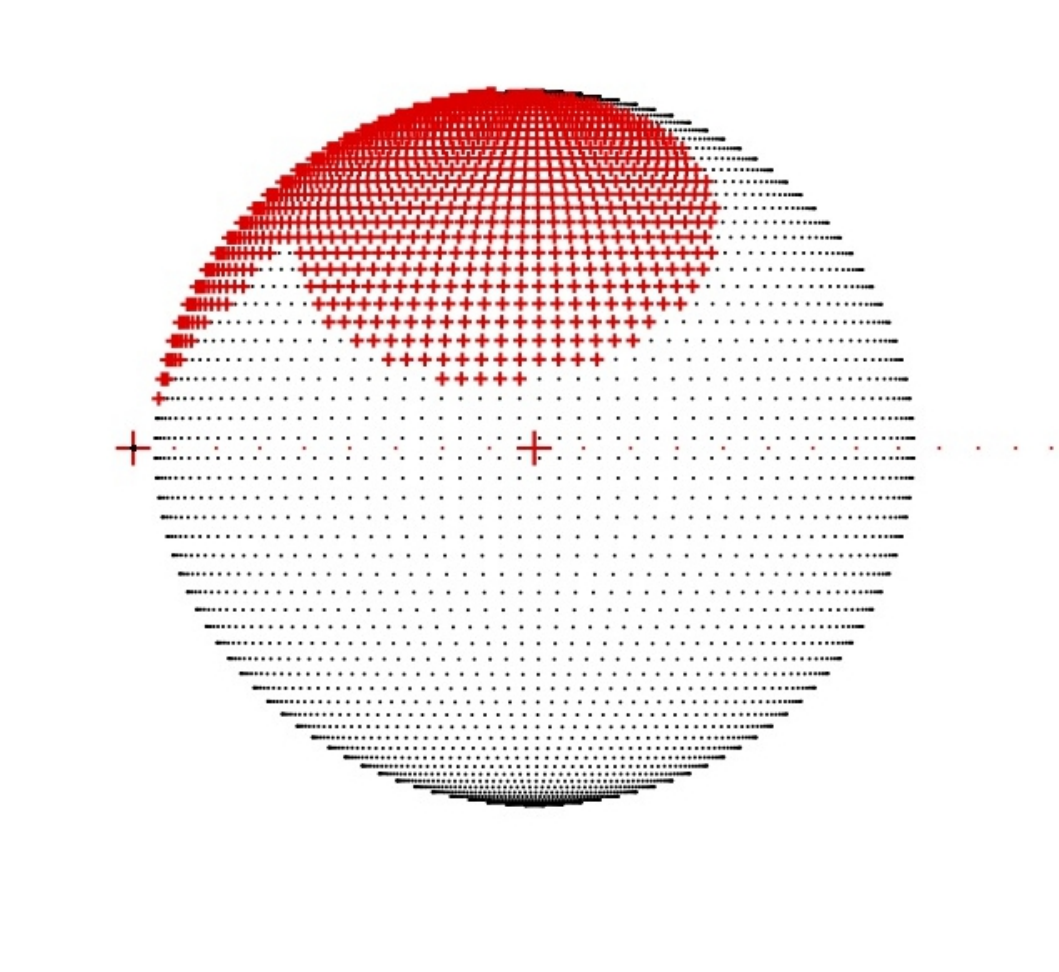} 
\includegraphics[width=30mm,height=30mm,angle=0,trim= 0 0 0 0]{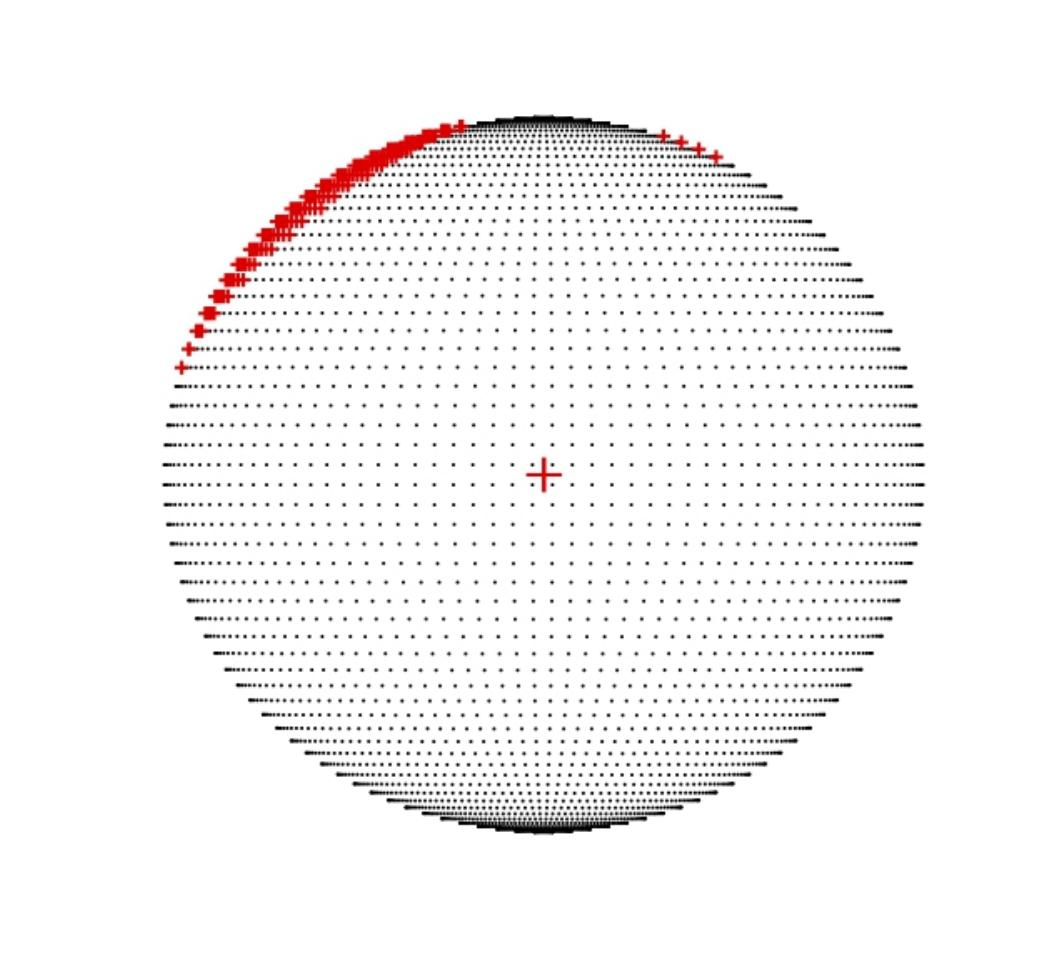} 
\includegraphics[width=30mm,height=30mm,angle=0,trim= 0 0 0 0]{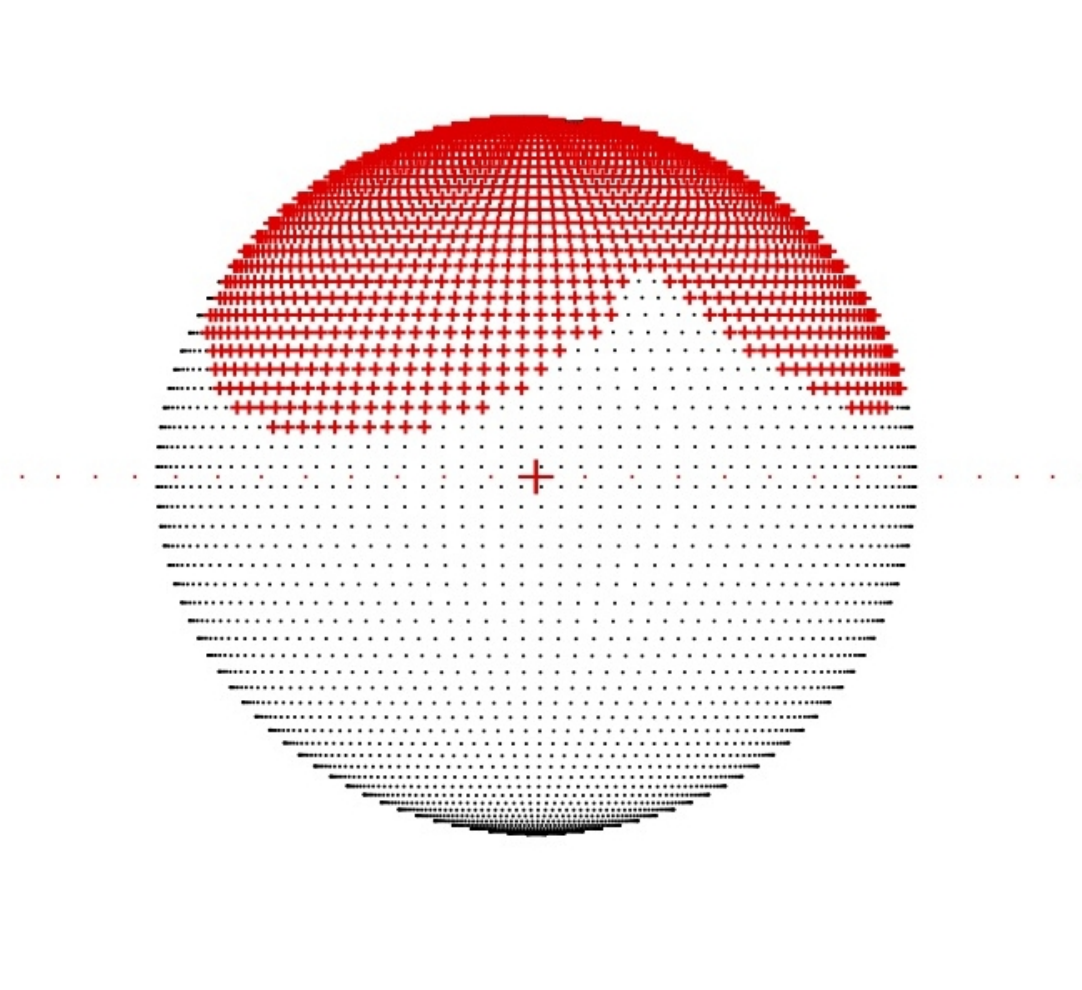} 
\end{minipage}
\caption{\label{spot} \it top panel: \rm Spot Model (solid lines) of the observed  V-, R-, and I-normalized flux (bullets, crosses, and squares respectively) versus rotation phase. An arbitrar y-axis offset of $-$0.2 and $-$0.4 has been added to the R and I fluxes, respectively. \it bottom panel: \rm 3D representation of the spotted star at the three different phases 0.2, 0.5, and  0.8.}
\vspace{0cm}
\end{figure*}

As mentioned earlier, we have indication that the radius of AG Tri A should be inflated   by 8\% \rm as consequence of a large covering fraction by spots. To make an estimate of the covering fraction by spots we performed a spot modeling of the data collected at the Zeta UMa Observatory, which exhibit the largest rotational modulation amplitude and are in three different photometric bands. 
We modeled the observed multi-band light curves using Binary Maker V 3.0 (\citealt{Bradstreet04}).  Binary Maker V 3.0 models are almost identical to those generated by Wilson-Devinney program (\citealt{Wilson71}) and uses   Roche equipotentials to create star surfaces.  In our modeling the second component is essentially ``turned off" (i.e., assigned a near zero mass and luminosity) in order to model a single rotating star. The gravity-darkening coefficient has been assumed $\nu$ = 0.25 (\citealt{Kopal59}), and  limb-darkening coefficients from \citet{Claret12} were also adopted. We adopted T$_{\rm eff}$ = 4334\,K, and inclination i = 90$^\circ$ as input parameters.  Thanks to the dependence of the light curve amplitude on the photometric band wavelength, we could constrain the spot temperature contrast and better determine the area of the spots  responsible for the flux rotational modulation. The temperature contrast between spots and surrounding photosphere  T$_{spot}$/T$_{phot}$ = 0.90 was found    by simultaneously modelling all three V, R, and I light curves, performing a number of iteration to finally find the best value that minimized the chi-squares in all three fits.   The only free parameters in 
our model remained the spot areas and the spot longitudes. The spots latitude cannot be constrained by photometry alone, especially in the case of an equator-on star as in our present case. \rm 
We found a satisfactory fit of all V, R, and I light curves using  two spots separated in longitude by  about 85$^{\circ}$ and with radius of about r = 40$\pm$5$^{\circ}$, which corresponds to a total covering fraction of about 25\% of the whole photosphere. Such a value refers to that component of spots unevenly distributed in longitude and, then, represents a lower value for the effective percentage of spotted surface. 
We stress that our aim is not to find a unique solution of the spot area and position, but to make a reasonable estimate of the spot area. Therefore, we just present one of the possible configurations.
 In the top panel of Fig.\,\ref{spot}, we plot the normalized flux in the  V, R, and I filters with overplotted the fits from our spot modeling. In the bottom panel of Fig.\,\ref{spot}, we plot a pictorial 3D model of the spotted star at three different rotation phases.\\
The large covering fraction by spots supports the inferred radius inflation invoked to conceal the measured photometric rotation period and projected rotational velocity of AG Tri A.

\section{Discussion}

Our system has two components with K4 and M1 spectral types corresponding, respectively, to masses  M$_{\rm A}$ = 0.85\,M$_\odot$ and  M$_{\rm B}$ = 0.35\,M$_\odot$ and with a projected
separation of 930\,AU.  Therefore, the mass difference between the two components
is about a factor 2. Differently than the mentioned cases (BD$-$21\,1074, TYC\,9300\,0891\,1AB/TYC\,9300\,0529\,1, and HIP\,10689/ HIP\,10679), this implies that in our subsequent discussion, also the mass difference may play a role in the observed rotation period difference between the two components.
The A component has a debris disc with mass M$_{dust}$ = 0.9 $\times$ 10$^{-3}$ M$_{\oplus}$, radius R$_{dust}$ = 8.9 AU, and T$_{dust}$ = 65\,K (\citealt{Riviere-Marichalar14}).
Neither disc nor infrared excess has been detected in the B component.
The A component has a rotation period P = 12.4\,d whereas the B component has a rotation period P = 4.66\,d. 
In their recent study, based on infrared surveys, \citet{Ribas14} show that the frequency of low-mass stars with discs 
decays exponentially with a decay time of $\tau$ = 2--3\,Myr for the inner discs, and $\tau$ = 4--6\, Myr for primordial discs. 
No primordial discs are detected at ages older than 8--10 Myr (\citealt{Jayawardhana99}, \citealt{Jayawardhana06}). 
Only a very low fraction of accretors (2--4\%) has been found in the 10 Myr old $\gamma$ Velorum association (\citealt{Frasca15}).
At the age of the $\beta$ Pictoris association, we expect that stars have their disc either completely accreted with no residual remnants or have debris discs.\\
In \citet{Messina14}, we have already presented the case of BD$-$21\,1074 AB, member of $\beta$ Pictoris Association, 
and consisting of two components, M1.5 + M2.5, that exhibit a large (70\%) rotation period difference, P$_{\rm A}$ = 9.3\,d and P$_{\rm B}$ = 5.4\,d.
Whereas the separation between the A and B components is 160\,AU, the B component has a nearby M5 companion C at 16\,AU.
We demonstrated that different initial rotation periods of the A and B components cannot account alone for the large 
difference of the present rotation periods. An enhanced disc dispersal of the component B, 
and a  consequent shortening of the disc-locking phase,
must be necessarily invoked. We attributed the enhanced dispersal to gravitational effects of C on the disc of B.\\
In the case of TYC9300-0891-1AB/TYC9300-0529-1 (\citealt{Messina16}), the perturber companion is at about 160\,AU and its effect on the disc lifetime seems
to have been marginal producing a much smaller (16\%) rotation period difference.
On the basis of this scenario, in the present case, the faster rotating component AG Tri B should have a nearby perturber that has shortened its disc lifetime 
allowing the B component to start spinning up earlier than the A component. However, high-contrast H-band imaging with the Subaru telescope by \citet{Brandt14}
 did not detect any companion candidates within 7$^{\prime\prime}$ ($\sim$ 300 AU projected) from both components. The achieved contrast limits range from
 $\Delta$H = 7.7\,mag at 0.25$^{\prime\prime}$ to $\Delta$H = 13.3\.,mag at 5$^{\prime\prime}$. If a companion exists, it must be either fainter than these
 limits or closer than 0.25$^{\prime\prime}$.\\
 To investigate the possible presence of a nearby companion to either AG Tri A or B, with the method of the radial velocities, we have analyzed 
the RV curves derived from the mentioned  high-resolution infrared spectra collected by  \citet{Bailey12}. We find that for both components the RV are variable at about the 4-$\sigma$ level.

 We have made use of the Generalized Lomb-Scargle periodogram analysis applied to the Keplerian periodogram (\citealt{Zechmeister09}) which is particularly suited to search for periodicities arising from Doppler effects that may be induced by a possible companion.


\begin{figure*}
\includegraphics[width=60mm,height=80mm,angle=90,trim= 0 0 0 0]{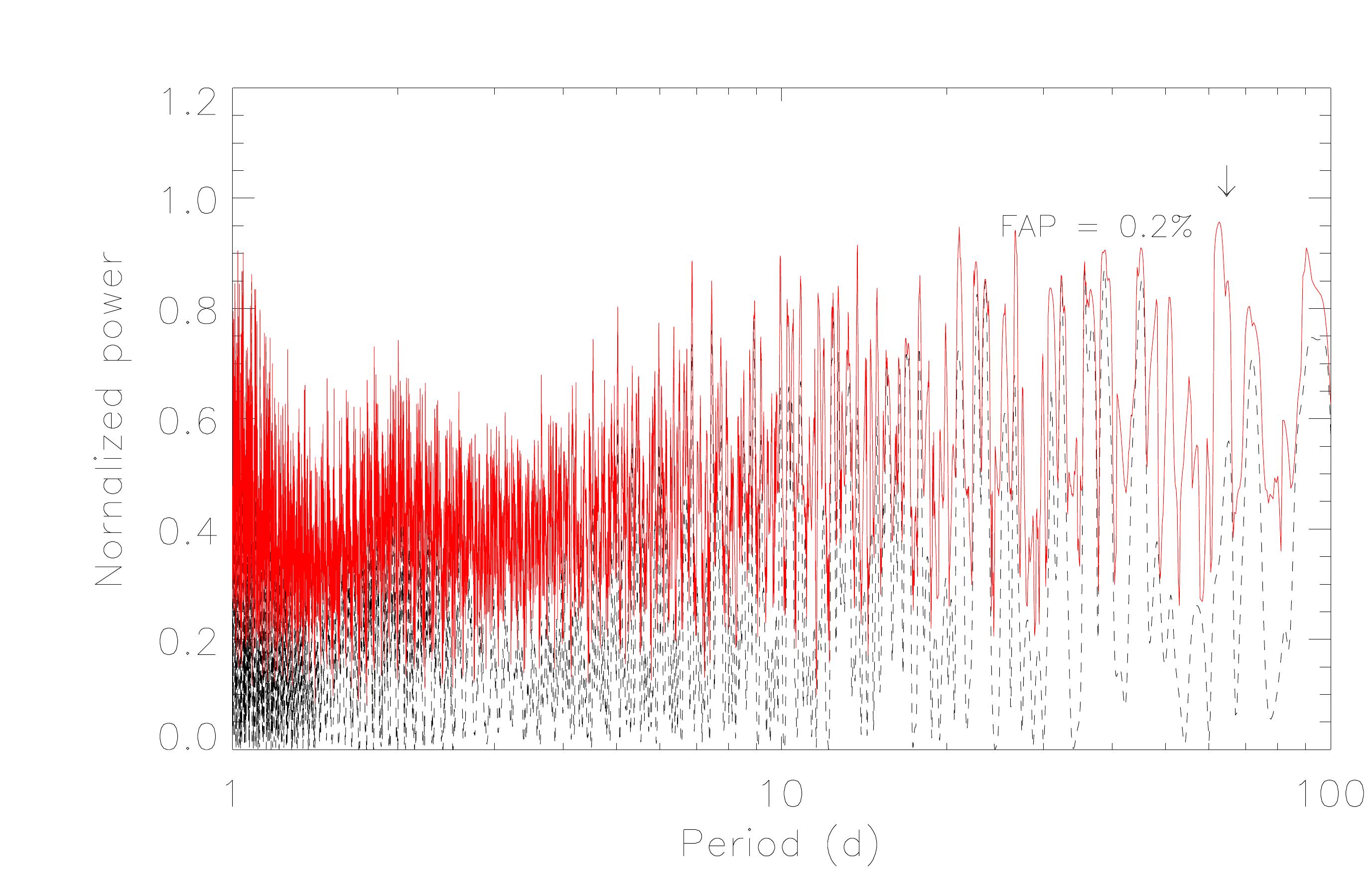}
\includegraphics[width=60mm,height=80mm,angle=90,trim= 0 0 0 0]{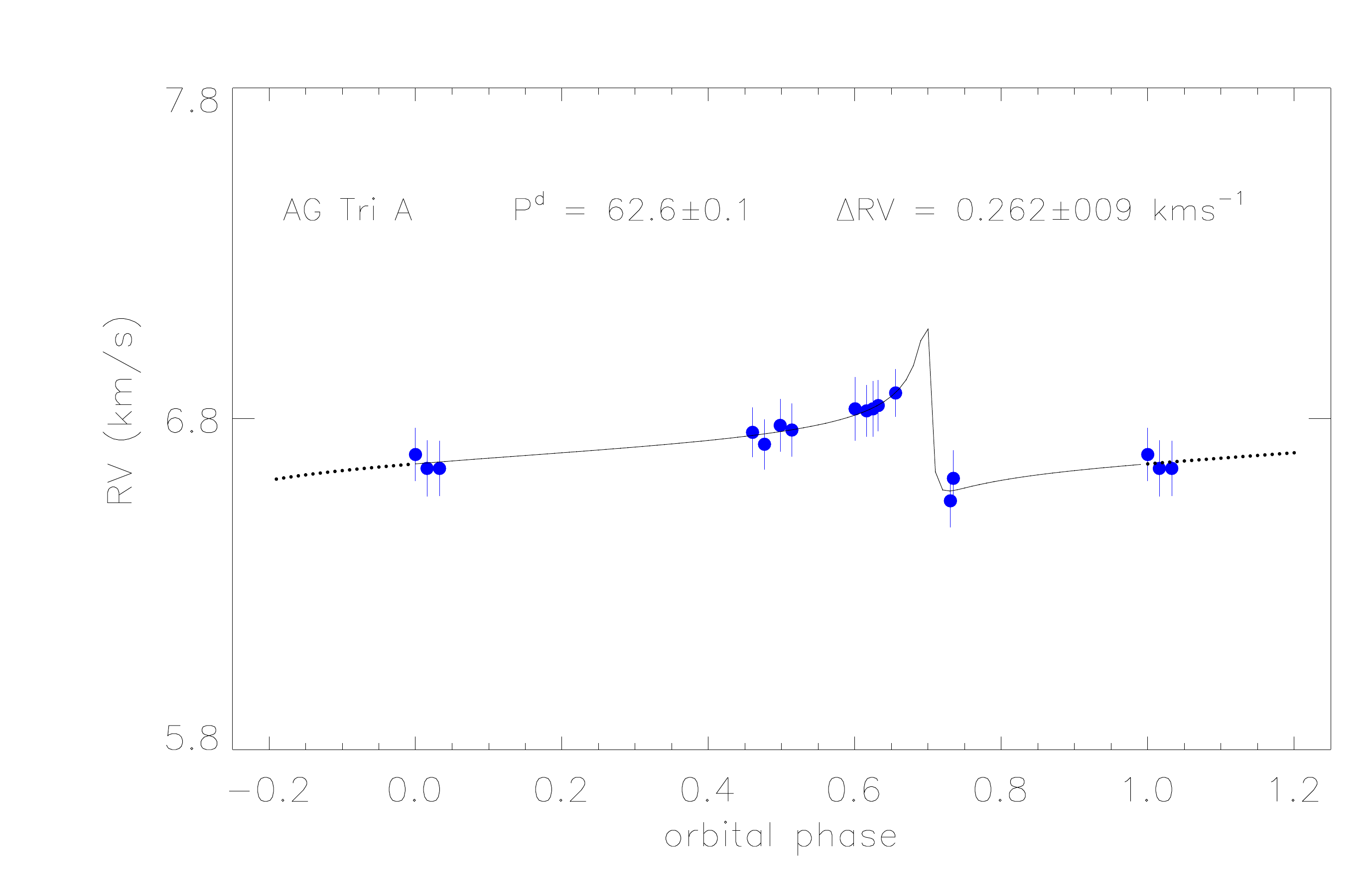}
\vspace{1cm}
\caption{\label{rv_period_a} \it left panel: \rm Lomb-Scargle (dashed line) and Generalized Lomb-Scargle (solid line) periodogram of the radial velocity measurements of AG Tri A collected by Bailey et al. (2012). The most significant periodicity (FAP = 0.2\%) is at P = 62.6\,d.
\it right panel: \rm radial velocity curve phased with the orbital period  P = 62.6\,d. The solid line is a Keplerian fit with amplitude $\Delta$(RV) = 256\,m\,s$^{-1}$ and eccentricity $e$ = 0.9. }
\end{figure*}

\begin{figure*}
\includegraphics[width=60mm,height=80mm,angle=90,trim= 0 0 0 0]{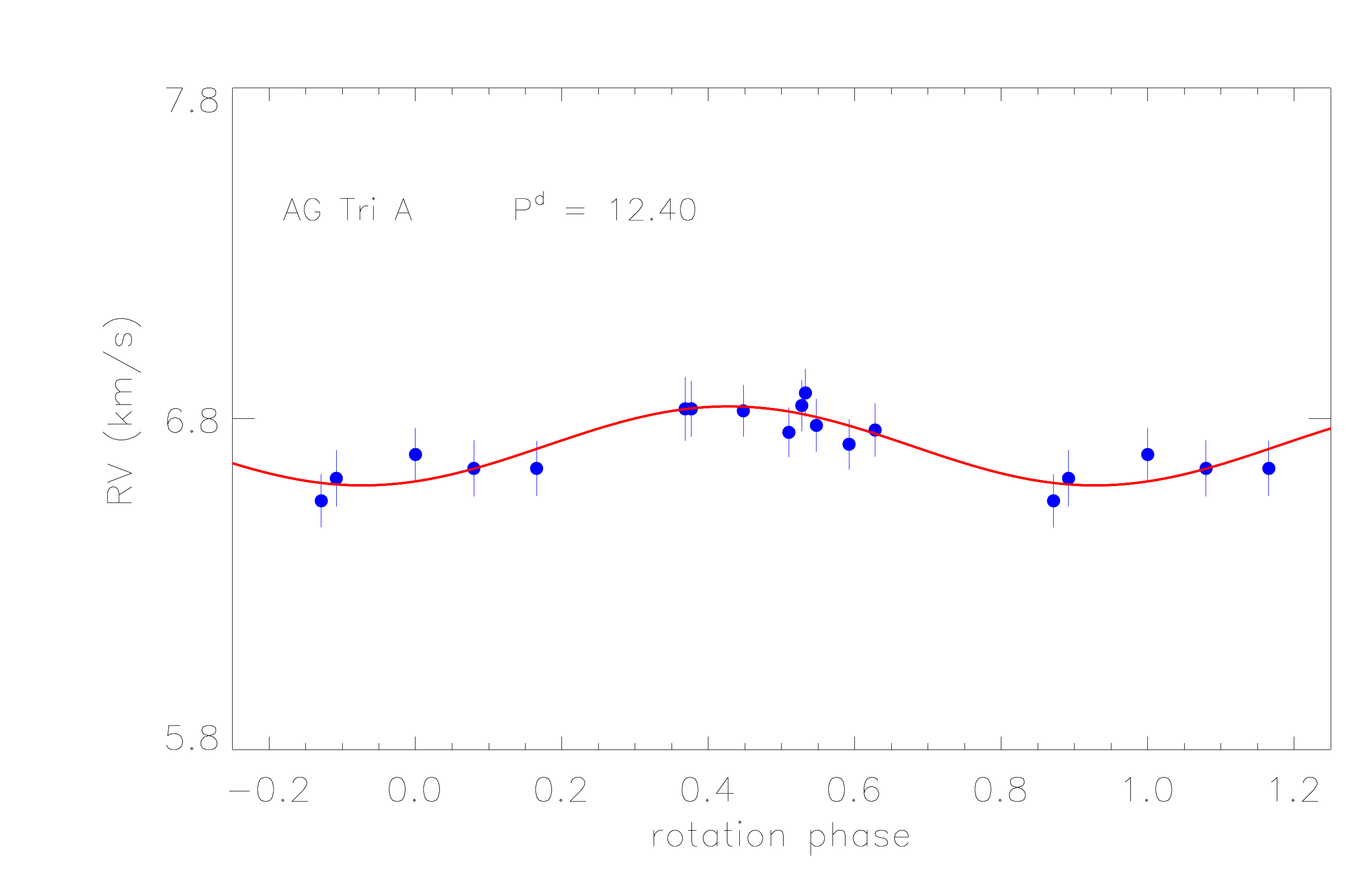}
\vspace{1cm}
\caption{\label{rv_period_a_bis} RV curve of AG Tri A phased with the rotation period P = 12.40\,d.}
\end{figure*}

In the case of the disc-less component AG Tri B, we did not detect any significant (FAP $<$ 1\%) periodicity.
In the case of AG Tri A, we find three highly significant periodicities in the range 10--70 days (see left panel of Fig.\,\ref{rv_period_a}), with the most significant (FAP = 0.2\%) corresponding to an orbital period P$_{orb}$ =  62.6$\pm$0.1\,d in a very eccentric orbit $e$ = 0.90  (see right panel of Fig.\,\ref{rv_period_a}).
However, we remind that detecting the presence of very low mass companions from RV curves in very active stars, as AG Tri A, is very challenging. Considering that we have 14 RV measurements that are very sparse, since they were collected in a total of four different runs over a total time span of about 2 years, and considering that  the most significant periodicity is accompanied by at least two other  peaks of comparable power, then we cannot take for granted about the Keplerian origin of the observed periodicity. 
	Indeed, if we phase the radial velocity data with the P = 12.4\,d rotation period we get (see Fig.\,\ref{rv_period_a_bis}) a very smooth RV curve with average residual of the same order of magnitude of the average RV uncertainty. \rm Rather, we tend to conclude with the currently available data that the measured variations are magnetic in origin and related to the magnetic field reconfiguration versus time.
Additional RV measurements are certainly needed to address the origin of the RV variations.   On the contrary, in the case of AG Tri B, when the RV data are phase with the P = 4.66\,d rotation period, no evidence of coherent rotational modulation  appears. \rm

On the basis of our analysis, the more likely scenario for the different rotation periods is that
the faster rotating component AG Tri B accreted its disc with typical (unperturbed) lifetime gaining all the angular momentum 
of its disc, whereas the slower rotating component AG Tri A received only a fraction of angular momentum from the disc, having formed the known debris disc,  remaining slower the star's rotation rate.
The present case seems to be  similar to the case of HIP\,10689/HIP\,10679 presented by \citet{Messina15}, 
which is a physical binary whose slower rotating component also hosts a debris disc.

\section{Conclusions}

We have carried out VRI photometric observations in two different observation runs 
of the two components A and B of the AG Tri
system in the young 23-Myr $\beta$ Pictoris stellar Association. We found that both components are variable 
in all filters and we were able to measure the stellar rotation periods of both components P$_{\rm A}$ = 12.4\,d and P$_{\rm B}$ = 4.66\,d.
The observed variability is coherent in all filters and exhibits an amplitude that decreases towards longer wavelengths.
This behavior is consistent with the presence of either hot or cool starspots as cause of the variability.
In the second run, the M1 component AG Tri B  showed a significant change in the spot configuration, from one major activity center
to two distinct active centers on opposite hemispheres.
We also detected on the M1 component AG Tri B  a flare event, the first one reported in the literature for this star, with an
increase of flux by 40\% with respect to the quiescent state. \\
Combining stellar radii, projected rotational velocities, and rotation periods, we found that AG Tri B has an inclination of the stellar rotation axis 
i $\sim$ 90$^{\circ}$. In the case of AG Tri A, to conceal projected rotational velocity and rotation period, we must invoke a radius inflation of   about 8\% \rm
produced by large covering fraction by starspots. Indeed, our spot modeling of the multi-band light curves of AG Tri A shows that this component has a minimum covering fraction by spots of 25\%. The high level of magnetic activity is likely the main cause
of the observed variation of the RV, which certainly deserves further investigation.\\
The presence of a debris disc around AG Tri A 
may be the main cause of the large difference in the rotation periods between the two components. More specifically, the disc could have prevented AG Tri A from gaining part of the angular momentum from the accreting disc, maintaining it rotation rate at slower regime with respect to the B component that accreted its disc completely.

\acknowledgments
The extensive use of the SIMBAD and ADS databases, 
operated by the CDS center, (Strasbourg, France), is  gratefully acknowledged. We acknowledge funding from the LabEx OSUG@2020 that allowed purchasing the ProLine PL230 CCD imaging system installed on the 1.25-m telescope at CrAO.
We give a special thank to the anonymous Referee for helpful comments that allowed us to improve the quality of the paper.

\end{document}